\documentstyle[12pt,epsfig]{article}

\newcommand{\ta}{\theta}
\newcommand{\e}{\epsilon}
\newcommand{\al}{\alpha}
\newcommand{\be}{\beta}
\newcommand{\sml}[1]{{\mbox{\scriptsize #1}}}
\newcommand{\bea}{\begin{eqnarray}}
\newcommand{\eea}{\end{eqnarray}}
\newcommand{\ind}{\hspace{2.0cm}}
\newcommand{\bd}[1]{{\bf #1}}
\newcommand{\half}{{\textstyle\frac{1}{2}}}
\newcommand{\tq}{{\textstyle\frac{3}{4}}}
\newcommand{\om}{\omega}
\newcommand{\ee}{e^+e^-}
\newcommand{\as}{\alpha_s}
\newcommand{\asPT}{\alpha_s^{\mbox{\scriptsize PT}}}
\newcommand{\eps}{\epsilon}
\newcommand{\epst}{\tilde{\eps}}
\newcommand{\beq}{\begin{equation}}
\newcommand{\eeq}{\end{equation}}
\newcommand{\cl}[1]{{\cal #1}}
\newcommand{\rf}[1]{(\ref{#1})}
\newcommand{\sect}[1]{\section{#1}\setcounter{equation}{0}
               \hspace{\parindent}\hspace{-0.14cm}}
\newcommand{\subsect}[1]{\subsection{#1}
               \hspace{\parindent}\hspace{-0.14cm}}
\newcommand{\nln}{\nonumber\\}
\newcommand{\Fldg}{F^{\mbox{\scriptsize leading}}}
\newcommand{\xp}{{\xi^\prime}}

\newcommand{\jhep}[3]{{\it J.~High Energy Phys.} {\bf #1} (#2) #3}
\newcommand{\npb}[3]{{\it Nucl. Phys.} {\bf B #1} (#2) #3}
\newcommand{\epj}[3]{{\it Eur. Phys.~J.} {\bf #1} (#2) #3}
\newcommand{\plb}[3]{{\it Phys. Lett.} {\bf B #1} (#2) #3}
\newcommand{\prd}[3]{{\it Phys. Rev.} {\bf D #1} (#2) #3}
\newcommand{\prlett}[3]{{\it Phys. Rev. Lett.} {\bf #1} (#2) #3}
\newcommand{\prep}[3]{{\it Phys. Rept.} {\bf #1} (#2) #3}
\newcommand{\ibid}[3]{{\it ibid.} {\bf #1} (#2) #3}

\begin{document}
\begin{titlepage}
\begin{flushright}
Bicocca-FT-01-05 \\
hep-ph/0105015 \\
May 2001
\end{flushright}              
\vspace*{\fill}
\begin{center}
{\Large \bf Power Corrections in Flavour-Singlet\\[1ex]
Deep Inelastic Scattering}
\end{center}
\par \vskip 5mm
\begin{center}
        G.E.~Smye\footnote{Research supported by the EU Fourth Framework
                Programme, `Training and Mobitily of Researchers', 
                Network `Quantum Chromodynamics and the Deep Structure of
                Elementary Particles, contract FMRX-CT98-0194 (DG12 - MIHT).}\\
        Dipartimento di Fisica, Universit\`{a} di Milano-Bicocca,\\
        and INFN Sezione di Milano, Italy.\\
\end{center}
\par \vskip 2mm
\begin{center} {\large \bf Abstract} \end{center}
\begin{quote}
We investigate the $1/Q^2$ power-suppressed corrections to structure functions in the flavour-singlet channel of deep inelastic lepton scattering arising from renormalon insertions into an initial-state gluon, as obtained using the dispersive approach. The pinch-technique is used as a convenient tool in the separation of contributions. 
\end{quote}
\vspace*{\fill}
\end{titlepage}

\sect{Introduction}
We have learned much about the strong interaction of particle physics from deep inelastic scattering (DIS) experiments over many years. This process is continuing, with present and future experiments generating more data over larger regions of phase space.

Alongside this ongoing experimental work, theoretical developments continue to be made. QCD, the established theory of strong interactions, continues to pose challenges: perturbation theory has been relatively successful at high energies, but the series expansion even here is at best asymptotic. In the non-perturbative r\'{e}gime lattice techniques are making advances. Yet a clean distinction between `perturbative' and `non-perturbative' cannot be made: all QCD observables involve some interplay between them.

Although the structure functions (for example) cannot be calculated using perturbative QCD, the general shape of their asymptotic $Q^2$ behaviour is well known: the observed Bjorken scaling at high $Q^2$ is violated by additional smaller terms. The dominant scaling violation is a logarithmic $Q^2$ dependence, originating in the scale dependence of the parton density functions of the incident hadron. This can be calculated using perturbative QCD and used to measure the strong coupling $\as$.

In addition to the logarithmic scaling violations, there are known to be contributions behaving as inverse powers of the hard scale, i.e.~as $1/Q^n$. These include both corrections due to the non-zero hadron mass $M$, which are suppressed by a factor $M^2/Q^2$, and non-perturbative power-suppressed terms arising from higher-twist operators in the operator product expansion. Such contributions are not included in fixed-order perturbative calculations, yet they are known to be important over the wide $Q^2$ range of available data.

Over a number of years the higher-twist power-suppressed terms of a wide variety of observables have been estimated using two related approaches, the `renormalon' (see \cite{ben} for a review) and `dispersive' \cite{dmw} models, although it is the dispersive approach that is used here. If we consider graphs with an arbitrary number of loop insertions in a gluon propagator, we assume that we can reconstruct a well-defined effective strong coupling at the scale of the gluon virtuality. The difference between this `true' coupling and that reconstructed using fixed order perturbation theory gives rise to non-perturbative corrections, which are typically power-behaved. For a given observable, the shape of the leading correction can be found; an estimation of its magnitude requires the additional assumption of universality. Thus, starting from perturbative QCD, we aim to investigate the transition to the non-perturbative region.

These approaches have been applied to various QCD observables, with the assumption of universality approximately holding \cite{univ}. In flavour non-singlet DIS there are results for structure functions \cite{dwdisstr, nonssf}, fragmentation functions \cite{nonsff} and event shape variables \cite{nonses}, while studies have been made of power corrections to structure functions \cite{stein,smye} and fragmentation functions \cite{smye2} in the flavour singlet channel.

\begin{figure}[ht]
\begin{center}
\epsfig{file=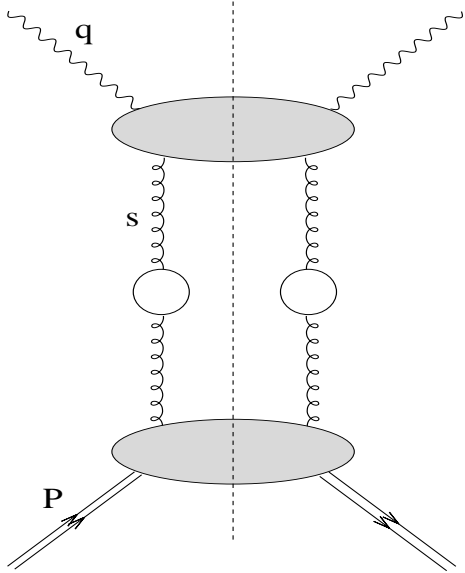,height=3.0in,width=2.4in}
\caption{\label{decompfig}Power corrections in flavour singlet DIS}
\end{center}
\end{figure}

The flavour singlet contribution to DIS is that involving the interaction of the gluons within the hadron. (We do not study the fermionic singlet, arising from the total sum of quark distributions, since these behave just like the non-singlet contribution \cite{dwdisstr, nonssf}.) Power corrections are calculated using the renormalon or dispersive models by making insertions into the gluon propagators as shown in figure \ref{decompfig}. Here the lower part of the diagram represents the finding of a virtual gluon within the proton, and the upper part represents photon-gluon fusion.

In the normal perturbative treatment of DIS, the asymptotic freedom of QCD enables us to treat the initial-state partons as free particles confined within the nucleon; so in a singlet calculation we would start with a free gluon and convolute the perturbative result with the gluon distribution function $g(x)$. We cannot however do this in a calculation of power corrections, since the models we use consider modifications to the gluon propagator (loop insertions in the renormalon model, or, equivalently, a `mass' in the dispersive approach). We therefore consider the initial-state gluon to be generated by some perturbative mechanism, the simplest of which is by radiation from an on-shell parton.

We immediately encounter a very serious problem: if the gluon is not on shell, both the upper and lower parts of this diagram are gauge-dependent. Gauge independence is achieved only when we include all the elements of some subset of diagrams of a given order in $\as$, and this includes diagrams that cannot be separated into the two halves of figure \ref{decompfig}: diagrams which do not have analogous gluon propagators in which to make insertions or from which to take a scale for the running coupling.

This is precisely the same problem encountered in the two-loop calculations of power corrections to event shapes in $\ee$ annihilation, as discussed in \cite{cpar}, except that the gluon is now in a different channel. We might thus expect a similar solution: namely, the use of the pinch technique \cite{ptech} to generate the running coupling at the scale of the gluon virtuality \cite{watson} in squared diagrams with two exchanged gluons, and the remainder of diagrams contributing to the power correction via some as-yet unknown mechanism.

\begin{figure}[ht]
\begin{center}
\epsfig{file=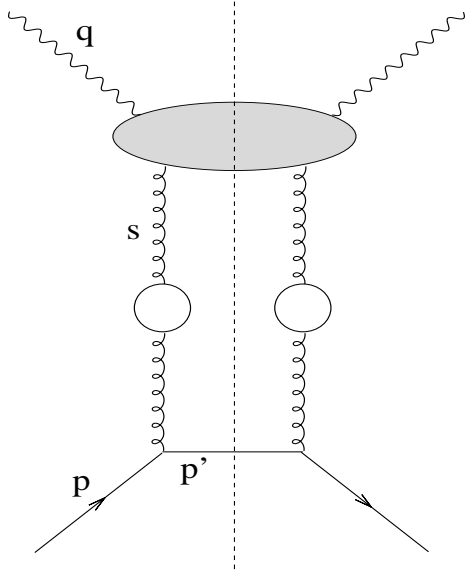,height=3.0in,width=2.4in}
\caption{\label{splitfig}Radiation of gluon from quark line}
\end{center}
\end{figure}

The first studies \cite{stein,smye,smye2} of these non-singlet quantities considered the incoming virtual gluon to be radiated from a quark line, as in figure \ref{splitfig}. This is the simplest case, since the lower half of the diagram is gauge independent, as is the upper half to leading order, $\cl{O}(\as)$.

There is however another problem to be overcome, concerning the interpretation of the lower half of the diagram. We may try to recover the singlet contribution to the power corrections by deconvoluting the full result with the quark-to-gluon splitting function, (this is the approach taken in \cite{stein}), or we may leave the result as it is and interpret it as a genuine second-order contribution. These two interpretations give very different predictions for the magnitudes of the power-suppressed corrections. I argue below (in section \ref{secglev}) that the latter approach should be adopted.

In section \ref{fsdis} we discuss the flavour singlet contribution to deep inelastic scattering. The dispersive approach to power corrections is then briefly reviewed in section \ref{disp}. Section \ref{prod} examines in detail the virtual gluon production represented by the bottom half of figure \ref{decompfig}, and the application of the pinch technique to this, while in section \ref{strfn} we calculate the contribution to the structure functions from photon-gluon fusion. Section \ref{secglev} investigates the distribution of virtual gluons in the proton, paying particular attention to the $1/Q^2$ power corrections. Finally a summary is given in section \ref{conc}.

\begin{figure}[ht]
\begin{center}
\epsfig{file=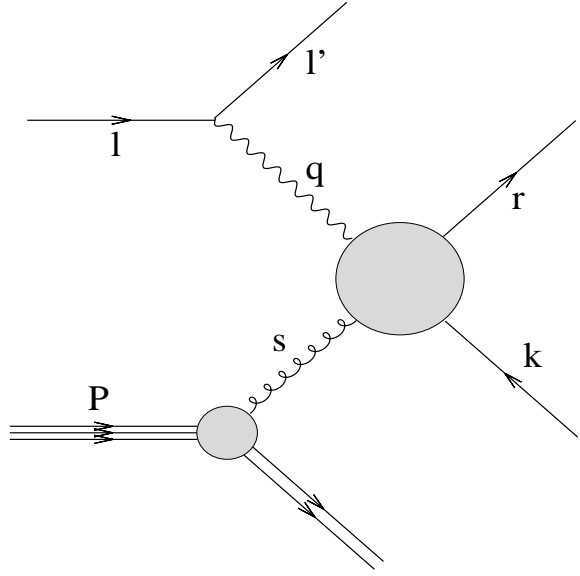, height=3.0in, width=3.0in}
\caption{\label{singdiag}Flavour singlet contribution to deep inelastic scattering}
\end{center}
\end{figure}

\sect{Flavour singlet DIS}
\label{fsdis}
Consider the deep inelastic scattering of a lepton with 4-momentum $l$ from a hadron with 4-momentum $P$, as shown in figure \ref{singdiag}. If the momentum transfer is $q$, the usual kinematic variables are $Q^2 = -q^2$, the Bjorken variable $x = Q^2/2P\cdot q$, and $y = P\cdot q/P\cdot l\simeq Q^2/xs$, where $s$ is the square of the energy in the c.m. frame.

Then the differential cross section is
\beq
\frac{d^2\sigma}{dx\, dQ^2} = \frac{2\pi\alpha^2}{Q^4}\left\{[1+(1-y)^2]F_T(x) + 2(1-y)F_L(x)\right\}\;,
\eeq
where $F_T(x) = 2F_1(x)$ and $F_L(x) = F_2(x)/x-2F_1(x)$ are the transverse and longitudinal structure functions, which also have a weak $Q^2$ dependence which we do not show explicitly. (For simplicity we are neglecting any contribution from weak interactions, i.e.~$\mbox{Z}^0$ or $\mbox{W}^\pm$ exchange.)

We can consider the photon to interact with an asymptotically free parton, moving collinearly with the hadron with momentum $s^\mu = xP^\mu/\xi$ ($x\le\xi\le 1$). Then in the parton model, to order $\as^0$, we have
\beq
\label{bornsf}
F_T(x) = \sum_q e_q^2 [q(x) + \bar{q}(x)] \hspace{0.5in} F_L(x) = 0 \;,
\eeq
where $q(x)$ and $\bar{q}(x)$ are the quark and antiquark distributions in the target hadron. Thus at this level there is no contribution arising from the gluon distribution in the hadron.

The $\cl{O}(\as)$ contributions are most easily given as the distribution in the final-state variable $\eta = P\cdot r/P\cdot q$, ($0\le\eta\le 1$):
\beq
\frac{d}{d\eta}F_i(x) = \frac{\as}{2\pi}\sum_q e_q^2 \int_x^1\frac{d\xi}{\xi}\left\{C_FC_{i,q}(\xi,\eta)[q(x/\xi)+\bar{q}(x/\xi)]+T_RC_{i,g}(\xi,\eta)g(x/\xi)\right\}\;,
\eeq
where $g(x)$ is the gluon distribution in the target hadron, $C_F = 4/3$, $T_R = 1/2$ and the coefficient functions $C_{i,j}(\xi,\eta)$ \cite{pr} are
\begin{eqnarray}
C_{T,q}(\xi,\eta) &=& \frac{\xi^2+\eta^2}{(1-\xi)(1-\eta)}+2\xi\eta+2\\
C_{L,q}(\xi,\eta) &=& 4\xi\eta\\
C_{T,g}(\xi,\eta) &=& [\xi^2+(1-\xi)^2]\frac{\eta^2+(1-\eta)^2}{\eta(1-\eta)}\\
C_{L,g}(\xi,\eta) &=& 8\xi(1-\xi)\;.
\end{eqnarray}
An integration over the entire range of $\eta$ from 0 to 1 requires the implementation of a factorisation scheme to regulate the collinear divergences. In our calculations this will be effected by the introduction of a small gluon `mass'. In addition the coefficient functions $C_{i,q}(\xi,\eta)$ acquire contributions at $\eta=1$ from virtual gluon emission.

We are interested in the singlet contributions $C_{i,g}(\xi,\eta)$, corresponding to photon-gluon fusion; the $1/Q^2$ power corrections from the non-singlet components $C_{i,q}(\xi,\eta)$ were successfully analysed in \cite{dwdisstr, nonssf}.

\sect{The dispersive approach to power corrections}
\label{disp}
We assume that the QCD running coupling $\alpha_s(k^2)$ can be defined for all positive $k^2$, and that apart from a branch cut along the negative real axis there are no singularities in the complex plane. It follows that we may write the formal dispersion relation:
\beq
\label{disprel}
\as(k^2) = - \int_0^\infty\frac{d\mu^2}{\mu^2+k^2}\rho_s(\mu^2)\; ,
\eeq
where the `spectral function' $\rho_s$ represents the discontinuity across the cut:
\beq
\label{rhosdef}
\rho_s(\mu^2)=\frac{1}{2\pi i}\biggl\{\as(\mu^2 e^{i\pi})-\as(\mu^2 e^{-i\pi})\biggr\} = \frac{1}{2\pi i}\mbox{Disc }\as(-\mu^2)\;.
\eeq
To lowest order in perturbation theory we have $\rho_s(\mu^2)=-\frac{\be_0}{4\pi}\vert\as(-\mu^2)\vert^2$.

Non-perturbative effects at long distances are expected to give rise to a non-perturbative modification to the perturbatively-calculated strong coupling at low scales, $\delta\as(\mu^2) = \as(\mu^2) - \asPT(\mu^2)$, where $\asPT(\mu^2)$ is the perturbatively-calculated running coupling \cite{dmw}. Note that here $\asPT(\mu^2)$ refers only to the contribution to the running coupling from a fixed (next-to-leading) order perturbative calculation, and so is itself well-behaved down to low scales, without any divergences: the Landau pole appears when we include an arbitrary number of loop insertions in the propagator. Hence $\delta\as(\mu^2)$ is assumed to be well-defined for all positive $\mu^2$.

We now consider the calculation of some observable $F$ in an improved approximation which takes into account fixed-order contributions plus those higher-order terms that lead to the running of $\as$. As is well documented, for processes involving a single gluon, it is required to calculate the relevant contributions as though the gluon had a small mass $\mu^2=\eps Q^2$. Calculations of power corrections in $\ee$ annihilation, in non-singlet DIS, and in the Drell-Yan process, use a single such gluon. In these cases the $1/Q^n$ corrections are found to be proportional to
\beq
\label{Andef}
\cl{A}_n \equiv \frac{C_F}{2\pi}\int_0^\infty \frac{d\mu^2}{\mu^2}\,\mu^n\,\delta\as(\mu^2)\;.
\eeq
Numerical values for these parameters must be obtained from data: fits of the $1/Q^2$ corrections to DIS structure functions suggest that $\cl{A}_2 \simeq 0.2 \mbox{ GeV}^2$ \cite{dwdisstr}.

However, DIS in the singlet channel involves two such gluons. Both gluons have an associated dispersive variable, so we obtain a characteristic function $\cl{F}(\eps_1,\eps_2)$, where $\eps_i = \mu_i^2/Q^2$, which is simply the observable $F$ calculated as though the gluons had masses $\mu_1$ and $\mu_2$, and without any factors of $\as$.

Since both gluons are constrained to have the same 4-momentum $s$, we can simplify this to require only one dispersive variable. By defining $\rho = - s^2 / Q^2$, we see that the dependence of the characteristic function $\cl{F}$ on $\eps_1$ and $\eps_2$ is given by
\beq
\label{Feq}
\cl{F}(\eps_1,\eps_2) = \int\frac{d\rho\,\rho f(\rho)}{(\rho+\eps_1)(\rho+\eps_2)}\;,
\eeq
where the integration limits and the function $f$ depend on the particular calculation. This may be re-expressed in the form
\beq
\label{Fdecomp}
\cl{F}(\eps_1,\eps_2) = \frac{\eps_1\hat{\cl{F}}(\eps_1)-\eps_2\hat{\cl{F}}(\eps_2)}{\eps_1-\eps_2}\;,
\eeq
where
\beq
\label{Fdec2}
\hat{\cl{F}}(\eps) = \int\frac{d\rho}{(\rho+\eps)}f(\rho)\;.
\eeq
So it is sufficient to perform the calculation with one `mass' set equal to zero, the other giving us the form of the characteristic function. 

The alternative expression
\beq
\hat{\cl{F}}(\eps) = \frac{1}{\eps}\int d\rho\,f(\rho)-\frac{1}{\eps}\int\frac{d\rho\,\rho f(\rho)}{\rho+\eps}
\eeq
shows how this relates to the slightly different definition of $\hat{\cl{F}}$ found in \cite{smye,smye2}. In these references it is shown that a characteristic function of the form \rf{Fdecomp} leads to a non-perturbative correction to the observable $F$ given by
\beq
\label{delF}
\delta F = \int_0^\infty\frac{d\mu^2}{\mu^2}\left(2\as(\mu^2)\delta\as(\mu^2)-[\delta\as(\mu^2)]^2\right)\hat\cl{G}(\mu^2/Q^2)\;,
\eeq
where, using Cauchy's theorem,
\beq
\hat{\cl{G}}(\eps) = -\frac{1}{2\pi i}\mbox{Disc}\hat{\cl{F}}(-\eps) = -\frac{1}{2\pi i}\biggl\{\hat{\cl{F}}(\eps e^{i\pi})-\hat{\cl{F}}(\eps e^{-i\pi})\biggr\} = f(\eps)\;.
\eeq
(Another method of arriving at \rf{delF} in this particular case is simply to insert the running coupling $\as(-s^2)$ into the matrix element before integrating.)

Since $\delta\as(\mu^2)$ is small in the perturbative r\'{e}gime, and vanishes as $\mu^2\to\infty$, the correction \rf{delF} depends on the behaviour of $\hat{\cl{F}}$ at small $\eps$. So we perform an expansion of this function about $\eps=0$.

Any divergent term in $\hat{\cl{F}}(\eps)$ is subtracted off. These are the terms causing the running of parton distributions, which gives rise to logarithmic scaling violations: the gluon mass here behaves as a regulator.

The remaining terms, which vanish as $\eps\to 0$, give power corrections. We will find the dominant terms to be:
\beq
\hat{\cl{F}}\sim a_1 \eps\log\eps
\qquad\Longrightarrow\qquad
\delta F = a_1 \frac{D_1}{Q^2}\;,
\eeq
and
\beq
\hat{\cl{F}}\sim \half a_2 \eps\log^2\eps
\qquad\Longrightarrow\qquad
\delta F = a_2\frac{D_1}{Q^2}\log\frac{D_2}{Q^2}\;,
\eeq
where the non-perturbative parameters $D_1$ and $D_2$ are defined by:
\begin{eqnarray}
\label{D1def}
D_1 &\equiv& \int_0^\infty \frac{d\mu^2}{\mu^2}\,\mu^2\,\left(2\as(\mu^2)\delta\as(\mu^2)-[\delta\as(\mu^2)]^2\right)\;,\\
\label{D2def}
\log D_2 &\equiv& \frac{1}{D_1}\int_0^\infty \frac{d\mu^2}{\mu^2}\,\mu^2\log\mu^2\,\left(2\as(\mu^2)\delta\as(\mu^2)-[\delta\as(\mu^2)]^2\right)\;.
\end{eqnarray}

While we expect the form of $\as(\mu^2)$, and hence $D_1$ and $D_2$, to be universal, we have as yet no numerical values for them. It will be necessary therefore to extract values for $D_1$ and $D_2$, either from experimental results or from some model of the form of $\alpha_s(\mu^2)$.

\section{Production of the gluon}
\setcounter{equation}{0}
\label{prod}
\subsect{Gluon kinematics and polarisations}
\label{kin}
Let us consider the radiation of a virtual gluon with 4-momentum $s$
from a massless parton with 4-momentum $p$, as shown for the case of a
quark in figure \ref{splitfig}. The parton model assumption is $p=xP/\xi$,
where $P$ is the 4-momentum of the incoming hadron. 
Since the gluon is virtual, it need not be collinear with the proton, nor 
need it be transversely polarised.

We describe the kinematics by three additional variables, $\om$
and $\xp$ defined by
\beq
\frac{\xp}{1+\om} = \frac{Q^2}{2s\cdot q}\hspace{1.0in}
\frac{\om}{\xp^2} = -\frac{s^2}{Q^2} = \rho\;,
\eeq
and $\theta$ the azimuthal angle in the Breit frame. Then the relevant
4-momenta are, in the Breit frame,
\begin{eqnarray}
q &=& \half Q(0,0,0,2)\label{qmom} \\ p &=& \half Q(1/\xi,0,0,-1/\xi)
\\ s &=& \half
Q(s_0,s_\perp\cos\theta,s_\perp\sin\theta,s_3)\;,\label{smom}
\end{eqnarray}
where
\begin{eqnarray}
s_0 &=& \frac{1+(1-2\xi/\xp)\om}{\xp} \\
\label{sperp}
s_\perp^2 &=& \frac{4\om(1-\xi/\xp)(1-\om\xi/\xp)}{\xp^2} \\ s_3 &=&
-\frac{1+\om}{\xp}\;.
\end{eqnarray}
The gluon is therefore produced with three degrees of freedom: $\xp$
determines the longitudinal momentum fraction $x/\xp$ of the gluon,
$\om$ is proportional to the gluon's virtuality, and $\theta$ gives
its azimuthal angle. If we are concerned only with the produced gluon,
and not with the parton that emitted it, then $\xi$ is also free. Note
that in the limit $\om\to 0$, the gluon 4-momentum $s$ becomes
$xP/\xp$, the usual parton model result, independent of $\xi$ 
and $\theta$. This limit gives the
standard photon-gluon fusion contribution to the observable under
consideration. However in section \ref{disp} we saw that it is terms
of higher order in $\rho$ that generate power corrections: the
transverse momentum of the gluon is crucially important. Consequently
we might like the gluon to have a 4-dimensional 4-momentum
distribution within the proton: this will be discussed
in section \ref{secglev}.

With these definitions, the integration measure for the splitting
becomes
\beq
\label{splmes}
\int\frac{d^3\bd{p}^\prime}{(2\pi)^3 2p^{\prime 0}} = \frac{1}{(4\pi)^2}\int_\xi^1\frac{d\xp}{\xp}\int_0^{2\pi}\frac{d\theta}{2\pi}\int_0^\xp\frac{d\om}{\om}(1-\om)(-s^2)\frac{\xi}{\xp}\;.
\eeq

Let us also introduce polarisation vectors $\eps_i^\mu$, ($i=1,2,3$),
for the radiated gluon. We must choose the $\eps_i^\mu$ to satisfy
$s\cdot\eps_i = 0$ and $\eps_i\cdot\eps_j = -\delta_{ij}$. Then we
have the identity
\beq
\label{epssum}
\sum_{i=1}^3\eps_i^\mu\eps_i^\nu = -g^{\mu\nu}+\frac{s^\mu s^\nu}{s^2}\;.
\eeq
For those contributions to figure \ref{decompfig} where both the top
and bottom halves are gauge-independent (such as the lowest-order
contribution to figure \ref{splitfig}), we can use \rf{epssum} to
replace the gluon propagators with sums over polarisation vectors,
thereby detaching the top and bottom halves of the diagram. Where it
is not gauge independent, this gives the contribution in Landau gauge.

For the purpose of later discussion, let us introduce two such sets of
polarisation vectors. Firstly, let us define the `natural'
polarisation vectors $\eps_i$ by imposing $q\cdot\eps_2 = q\cdot\eps_3
= p\cdot\eps_3 = 0$. Then the identity \rf{epssum} gives us
\begin{eqnarray}
\label{nat}
\eps_1^\mu &=& \frac{(1+\om)\xp\,s^\mu+2\om\,q^\mu}{iQ(1-\om)\sqrt{\om}} \nln
\eps_2^\mu &=& \frac{[(1+\om)\xp-2\om\xi]\,s^\mu+[2-(1+\om)\xi/\xp]\om\,q^\mu-(1-\om)^2\xi\,p^\mu}{Q(1-\om)\sqrt{\om(1-\xi/\xp)(1-\om\xi/\xp)}}
\end{eqnarray}
and $\eps_3$ is orthogonal to the vector space spanned by $s$, $q$ and
$p$.

The vectors $\eps_2$ and $\eps_3$ are both orthogonal to $q$, and so
are the transverse polarisation vectors in frames in which $\bd{s}$ and
$\bd{q}$ are (anti-)parallel, and thus in all physical frames.
Also, since $s$ is spacelike, the
longitudinal polarisation vector $\eps_1$ is timelike. (The factor $i$
in the expression for $\eps_1$ then causes the identity \rf{epssum} to
hold as written; we could alternatively have introduced a factor $-1$
in the sum over polarisation vectors.)

These are the natural polarisation vectors because $\eps_1$ is
longitudinal and $\eps_2$ and $\eps_3$ are transverse: they are thus
the polarisation vectors that should be used when considering parton
splitting, and the splitting functions arise naturally from their use.

Secondly, let us define the `diagonal' basis of polarisation vectors
$\epst_i$ by imposing $p\cdot\epst_2 = p\cdot\epst_3 = q\cdot\epst_3 =
0$. It will be seen below that this is the basis that diagonalises the
matrix describing gluon production, and is formed by mixing
$\eps_1^\mu$ and $\eps_2^\mu$ to give the new basis vectors
\begin{eqnarray}
\label{diag}
\epst_1^\mu &=& \frac{2\xp\,p^\mu-\xp\,s^\mu}{iQ\sqrt{\om}} \nln
\epst_2^\mu &=& \frac{[2\xp-(1+\om)\xi]\,p^\mu-\xp s^\mu-(\om\xi/\xp)\,q^\mu}{Q\sqrt{\om(1-\xi/\xp)(1-\om\xi/\xp)}} \nln
\epst_3^\mu &=& \eps_3^\mu\;.
\end{eqnarray}
The disadvantage of this basis is that $\epst_1$ and $\epst_2$ are
neither transverse nor longitudinal, and therefore are not suitable
for a discussion of parton splitting or of splitting functions. The
use of this basis will be denoted by a tilde on appropriate symbols.

\subsect{Radiation from quark line}
The matrix element for the process shown in figure \ref{splitfig},
integrated using the measure \rf{splmes}, gives the following
parton-level contribution to some observable $F$:
\beq
\label{split}
F^q = \frac{\as
C_F}{2\pi}\sum_{i,j}\int_\xi^1\frac{d\xp}{\xp}\int_0^{2\pi}\frac{d\theta}{2\pi}\int_0^\xp
\frac{d\om}{\om}(1-\om)\frac{\xi}{\xp}\left[\half\delta_{ij}-\frac{2(p\cdot\eps_i)(p\cdot\eps_j)}{s^2}\right]F_{ij}(\xi,\xp,\om,\theta)\;,
\eeq
where $F_{ij}$ is the contribution due to photon-gluon fusion with the
appropriate gluon polarisations.

Using the natural basis \rf{nat}, we find
\beq
\label{split2}
F^q = \frac{\as
C_F}{2\pi}\sum_{i,j}\int_\xi^1\frac{d\xp}{\xp}\int_0^{2\pi}\frac{d\theta}{2\pi}\int_0^\xp\frac{d\om}{\om}\;M_{ij}^q(\xi/\xp,\om)F_{ij}(\xi,\xp,\om,\theta)\;,
\eeq
where
\beq
\label{split3}
M_{ij}^q(z,\om) = \left(\begin{array}{ccc}\frac{-2(1-z)(1-\om
z)}{(1-\om)z} & \frac{(2-z-\om z)\sqrt{(1-z)(1-\om z)}}{i(1-\om)z}
& 0 \\[0.1in]
\frac{(2-z-\om z)\sqrt{(1-z)(1-\om z)}}{i(1-\om)z} & \frac{(2-z-\om z)^2}{2(1-\om)z} & 0 \\[0.1in]
0 & 0 & \frac{(1-\om)z}{2} \end{array}\right)_{ij}
\eeq

Several comments may be made about this matrix:

\begin{enumerate}
\item The elements on the diagonal correspond to the exchange of a gluon with definite polarisation; the two off-diagonal elements give rise to an interference term between the longitudinal and one of the transverse polarisations.

\item The usual $q\to g$ splitting function can be recovered from the total transverse piece $M_{22}^q+M_{33}^q$ in the limit $\om\to 0$:
\beq
\lim_{\om\to 0}\left(M_{22}^q+M_{33}^q\right) = P_{q\to g}(z) = \frac{1+(1-z)^2}{z}\;.
\eeq
The standard factorised expression for the leading (logarithmic) divergence
arises from the fact that as $\om\to 0$, $s^\mu\to xP^\mu/\xp$ and $\e_1^\mu\sim s^\mu/\sqrt{s^2}$, so $F_{11}$, $F_{12}$ and $F_{21}$ all become subleading and (for an unpolarised observable) $F_{22}$ and $F_{33}$ become equal and functions of $\xp$ only. Thus the leading logarithmic divergence in $F$ is
\beq
\Fldg(\xi) = \frac{\as C_F}{2\pi}\int_\xi^1\frac{d\xp}{\xp}\int_0^\xp\frac{d\om}{\om}P_{q\to g}(\xi/\xp)C_g(\xp)
\eeq
where $C_g$ represents the contribution from a transversely-polarised
initial-state gluon.

This however applies only to the leading divergence: other pieces, in
particular those non-divergent pieces generating power corrections, do 
not factorise in
this way. In general, the functions $F_{ij}$ have dependence on $\xi$
and $\theta$ as well as $\om$ and $\xp$, i.e.~for the purposes of
higher twist contributions the gluon `remembers' the momentum fraction
of the particle that emitted it.

\item The matrix $M_{ij}^q$ has zero determinant, and hence has a zero eigenvalue. The other two eigenvalues are both equal to $(1-\om)z/2$. It is diagonalised if we use the diagonal basis \rf{diag}, in which case the decomposition \rf{split} simplifies considerably to:
\beq
F^q = \frac{\as
C_F}{2\pi}\int_\xi^1\frac{d\xp}{\xp}\int_0^{2\pi}\frac{d\theta}{2\pi}\int_0^\xp\frac{d\om}{\om}\frac{(1-\om)\xi}{2\xp}\left\{\tilde{F}_{22}+\tilde{F}_{33}\right\}\;.
\eeq

In other words, only two gluon polarisations, $\epst_2^\mu$ and
$\epst_3^\mu$, are permitted, but they are not the transverse
polarisations assumed in the parton model. Since this basis fails to
make apparent the form of the leading divergence and
does not naturally give rise to the splitting function, it is clear
that any attempt to remove the initial quark line from the power
corrections calculated in \cite{smye} and \cite{smye2}
 cannot be simply a
matter, as with the logarithmic divergences, of deconvoluting with the
quark to gluon splitting function. Also, since all the elements of
\rf{split3} contribute to the power correction we would need to be
able to include the unphysical polarisations in the gluon
distribution.
\end{enumerate}

\begin{figure}[ht]
\begin{center}
\epsfig{file=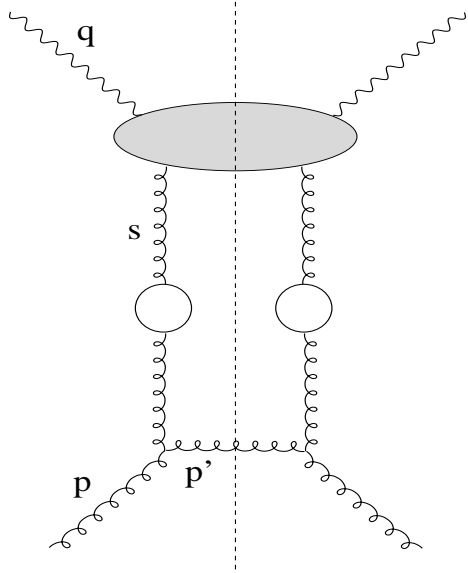,height=3.0in,width=2.4in}
\caption{\label{spglufig}Radiation of gluon from gluon line}
\end{center}
\end{figure}

\subsect{Radiation from gluon line}
The radiation of the virtual gluon from a gluon line is shown in
figure \ref{spglufig}. This diagram is gauge-dependent, and while its 
logarithmically divergent piece is gauge-independent and given by the 
$g\to g$ splitting function, the remainder of the diagram, and in 
particular the pieces generating power corrections, are not.

The matrix element for the process shown in figure \ref{spglufig},
integrated using \rf{splmes} and using the natural basis of polarisation 
vectors \rf{nat}, gives the following contribution to the observable $F$:
\beq
\label{glu2}
F^g = \frac{\as
C_A}{2\pi}\sum_{i,j}\int_\xi^1\frac{d\xp}{\xp}\int_0^{2\pi}\frac{d\theta}{2\pi}\int_0^\xp\frac{d\om}{\om}\;M_{ij}^g(\xi/\xp,\om)F_{ij}(\xi,\xp,\om,\theta)\;,
\eeq
where
\beq
\label{glu}
M_{ij}^g(z,\om) = \left(\begin{array}{ccc}\frac{-(2-z)^2}{2z} &
\frac{(2-z)(2-2z+z^2)}{2iz\sqrt{1-z}} & 0 \\[0.1in]
\frac{(2-z)(2-2z+z^2)}{2iz\sqrt{1-z}} & \frac{2-4z+4z^2-2z^3+z^4}{z(1-z)} & 0 \\[0.1in]
0 & 0 & \frac{z(2-2z+z^2)}{1-z} \end{array}\right)_{ij} +
\cl{O}(\om)\;.
\eeq

To compare with the quark case, several comments should be made:

\begin{enumerate}
\item The terms of $\cl{O}(\om)$ are gauge-dependent, and therefore of not much use until we can isolate pieces of other diagrams to give a gauge-invariant total. The pinch technique (see next section) succeeds in giving a gauge-invariant result, but at the expense of changing this leading order matrix.

\item Again there are on-diagonal elements corresponding to exchange of a gluon of definite polarisation, and off-diagonal elements giving interferences. In fact all six off-diagonal elements are non-zero at $\cl{O}(\om)$ level, unlike the quark case. In addition the use of the basis $\epst_i^\mu$ does not diagonalise the matrix, even to leading order.

\item The transverse elements $M_{22}^g$ and $M_{33}^g$ yield polarised splitting functions, and their sum gives
\beq
\lim_{\om\to 0}\left(M_{22}^g+M_{33}^g\right) = \frac{2(1-z+z^2)^2}{z(1-z)} = P_{g\to g}(z)\;.
\eeq
Thus the leading divergence is the well-known result
\beq
\Fldg(\xi) = \frac{\as C_A}{2\pi}\int_\xi^1\frac{d\xp}{\xp}\int_0^\xp\frac{d\om}{\om}P_{g\to g}(\xi/\xp)C_g(\xp)\;.
\eeq

\item The matrix $M_{ij}^g$ diverges at $z=1$, i.e.~when $\xp=\xi$. This is due to the gluon with momentum $p^\prime$ becoming soft, and the divergence is cancelled by virtual corrections.
\end{enumerate}

\subsect{Use of the pinch technique}
\label{pt}
When we considered the diagram in figure \ref{spglufig}, we found that
the terms giving rise to power corrections (i.e.~the terms in the
integrand of \rf{glu2} that do not diverge as $\om\to 0$) are
gauge-dependent. We must consequently include pieces of other diagrams
to restore gauge invariance. We could add in all $\cl{O}(\as^2)$
diagrams that contribute to the process $\gamma^* g\to q\bar{q}g$, as
in `Milan factor' calculations for event shape variables 
(see \cite{cpar}), but that would force us to include
diagrams unlike that of figure \ref{decompfig}: it is not clear what
relation (if any) these diagrams have to the scale chosen for the
running coupling.

\begin{figure}[ht]
\begin{center}
\epsfig{file=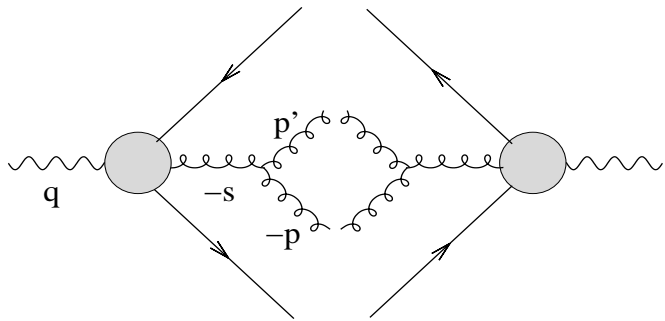,height=1.8in,width=3.6in}
\caption{\label{eefig}Diagram from $\ee$ annihilation related by crossing to the present case.}
\end{center}
\end{figure}

First we notice that figure \ref{spglufig} is related by crossing to the $\ee$ annihilation diagram shown in figure \ref{eefig}. So, just as in $\ee$ annihilation in \cite{cpar}, we can apply the pinch technique to the diagram to define a gauge-invariant contribution. But while the pinch technique has previously been applied to internal gluon loops \cite{watson} and cut loops containing outgoing partons \cite{cpar}, it is now being applied to a cut loop where one of the particles is incoming and the other outgoing. This difference is irrelevant since the algebra is identical. 

The matrix element for figure \ref{eefig}, on applying the pinch technique, is given in equation (A.9) of \cite{cpar}: in our notation it may be written
\beq
\vert M\vert^2 = g^2 C_A \frac{1}{s^4}\left\{8(s^2g_{\al\be}-s_\al s_\be)+2(p_\al^{}+p^\prime_\al)(p_\be^{}+p^\prime_\be)\right\}\sum_{i,j}\e_i^\al \e_j^\be F_{ij}(s)\;,
\eeq
where $F_{ij}$ is the contribution from the shaded blob. Using the crossing relation and integrating using \rf{splmes}, we obtain the following form for the observable $F$:
\beq
\label{glusp}
F^g = \frac{\as
C_A}{2\pi}\sum_{i,j}\int_\xi^1\frac{d\xp}{\xp}\int_0^{2\pi}\frac{d\theta}{2\pi}\int_0^\xp
\frac{d\om}{\om}(1-\om)\frac{\xi}{\xp}\left[2\delta_{ij}-\frac{2(p\cdot\eps_i)(p\cdot\eps_j)}{s^2}\right]F_{ij}(\xi,\xp,\om,\theta)\;.
\eeq
Note that this differs from the incoming quark case only by having a
different constant multiplying $\delta_{ij}$. In fact the only
linearly independent terms allowed inside the square bracket,
consistent with gauge invariance and having the correct dimensions,
are the two shown. This was also seen in the $\ee$ annihilation calculation
of \cite{cpar}.

Using the natural basis \rf{nat}, we find
\beq
\label{glusp2}
F^g = \frac{\as
C_A}{2\pi}\sum_{i,j}\int_\xi^1\frac{d\xp}{\xp}\int_0^{2\pi}\frac{d\theta}{2\pi}\int_0^\xp\frac{d\om}{\om}\;M_{ij}^g(\xi/\xp,\om)F_{ij}(\xi,\xp,\om,\theta)\;,
\eeq
where
\beq
\label{Mpt}
M_{ij}^g(z,\om) = \left(\begin{array}{ccc}\frac{-(2-3z+\om
z)(2+z-3\om z)}{2(1-\om)z} & \frac{(2-z-\om z)\sqrt{(1-z)(1-\om
z)}}{i(1-\om)z} & 0 \\[0.1in]
\frac{(2-z-\om z)\sqrt{(1-z)(1-\om z)}}{i(1-\om)z} & \frac{(2-z-\om z)^2}{2(1-\om)z}+\frac{3(1-\om)z}{2} & 0 \\[0.1in]
0 & 0 & 2(1-\om)z \end{array}\right)_{ij}
\eeq

As previously, there are several comments to be made:

\begin{enumerate}
\item Although this matrix is gauge invariant, it does not reduce to \rf{glu} in the limit $\om\to 0$. One consequence of this is that the $g\to g$ splitting function is no longer trivially recovered from the total transverse piece $M_{22}^g+M_{33}^g$ in the limit $\om\to 0$:
\beq
\label{ptsplfn}
\lim_{\om\to 0}\left(M_{22}^g+M_{33}^g\right) = \frac{2(1-z+2z^2)}{z} = P_{g\to g}(z) - \frac{2z^3}{1-z}\;,
\eeq
where the usual splitting function is
\beq
P_{g\to g}(z) = \frac{2(1-z+z^2)^2}{z(1-z)}\;.
\eeq

The additional piece arises from the fact that the pinch part of the diagram was removed (or alternatively that the pinch parts of other diagrams were added) in order to secure gauge invariance. This can be understood as follows: the factor $1/(1-z)$ in the splitting function can only arise from the collinear limit of the gauge-dependent piece $(n\cdot p)/(n\cdot p^\prime)$ in the sum over polarisations of the $p^\prime$ gluon in figure \ref{spglufig}. In order to achieve gauge invariance we must add or remove terms so as to cancel all $n$-dependence, not just in the collinear limit but identically for all $p$ and $p^\prime$, and so terms with the factor $1/(1-z)$ must necessarily disappear.

This happens even though the remaining diagrams themselves have no logarithmic divergence associated with the gluon splitting (although of course they may have divergences arising elsewhere): applying the pinch technique separates the diagrams into a pinch part and a remainder, which have equal and opposite logarithmically divergent pieces. Thus this is not a sensible way to study splitting functions. Also, since the only two linearly-independent gauge-invariant terms with the correct dimension are those in \rf{glusp}, and no linear combination of these can reproduce the splitting function, we see that we cannot have a fully gauge invariant expression for the splitting $g\to g+g^*$ (with some suitable definition for what is meant by $g^*$, e.g.~use of pinch technique) while retaining the full splitting function.

We also notice that the contribution \rf{ptsplfn} is no longer invariant under $z\to 1-z$: this is because the two daughters of the splitting are no longer identical, one being a real on-shell gluon and the other a virtual gluon with a modified propagator.

\item The difference between the production matrix employing the pinch technique and that using standard perturbation theory is
\beq
\delta M_{ij}^g(z,\om) = \left(\begin{array}{ccc} 2z & \frac{-(2-z)z}{2i\sqrt{1-z}} & 0 \\[0.1in]
\frac{-(2-z)z}{2i\sqrt{1-z}} & \frac{-z^3}{1-z} & 0 \\[0.1in]
0 & 0 & \frac{-z^3}{1-z} \end{array}\right)_{ij} + \cl{O}(\om)
\eeq
(where $\cl{O}(\om)$ terms are gauge-dependent). This is the contribution that gives the additions to the splitting function. It becomes zero as $z\to 0$, which is as the interacting virtual gluon becomes soft.

\item As in the quark and conventional gluon cases, there are both on- and off-diagonal elements. Now, however, the matrix $M_{ij}^g$ is diagonalised using the diagonal basis \rf{diag}, just as in the case where the emission is from a quark line. (Indeed one can trivially see that any linear combination of the two terms in \rf{glusp} is diagonalised using this basis.) One eigenvalue is $3(1-\om)z/2$, and the other two are both $2(1-\om)z$. Then the decomposition \rf{glusp} simplifies considerably to:
\beq
F^g = \frac{\as
C_A}{2\pi}\int_\xi^1\frac{d\xp}{\xp}\int_0^{2\pi}\frac{d\theta}{2\pi}\int_0^\xp\frac{d\om}{\om}\frac{2(1-\om)\xi}{\xp}\left\{\tq\tilde{F}_{11}+\tilde{F}_{22}+\tilde{F}_{33}\right\}\;.
\eeq
\end{enumerate}

\sect{Structure functions}
\label{strfn}
Consider now the part of the process involving the interaction of the
virtual gluon within the proton with the virtual photon. This is the
process shown in diagram \ref{singdiag}, which corresponds to the
quantities $F_{ij}$ above.

Working in the Breit frame, the 4-momentum of the incoming photon and
gluon are given by \rf{qmom} and \rf{smom}; in this part of the
process the variables $\xp$, $\om$ and $\theta$ are considered fixed. Let us
introduce the variables $\eta = P\cdot r/P\cdot q$, $\bar{\eta} =
P\cdot k/P\cdot q$ , $\chi$ the azimuthal angle between $r$ and $s$,
and $\bar{\chi}$ the azimuthal angle between $k$ and $s$. The
kinematics are then given by:
\begin{eqnarray}
r &=& \half Q
(z_0,z_\perp\cos(\chi+\theta),z_\perp\sin(\chi+\theta),z_3) \\ k &=&
\half Q
(\bar{z}_0,\bar{z}_\perp\cos(\bar{\chi}+\theta),\bar{z}_\perp\sin(\bar{\chi}+\theta),\bar{z}_3)\;.
\end{eqnarray}

The definitions of $\eta$ and $\bar{\eta}$ along with the on-shell
conditions for the outgoing particles require that
\begin{eqnarray}
z_0 = \eta+\frac{z_\perp^2}{4\eta}\hspace{0.75in} & & z_3 =
\eta-\frac{z_\perp^2}{4\eta} \\
\bar{z}_0 = \bar{\eta}+\frac{\bar{z}_\perp^2}{4\bar{\eta}}\hspace{0.75in} & &
\bar{z}_3 = \bar{\eta}-\frac{\bar{z}_\perp^2}{4\bar{\eta}}\;,
\end{eqnarray}
whence conservation of the 0th and 3rd components of 4-momentum give the
conditions
\begin{eqnarray}
\eta + \bar{\eta} &=& 1-\om\xi/\xp^2 \\
\frac{z_\perp^2}{4\eta}+\frac{\bar{z}_\perp^2}{4\bar{\eta}} &=& \frac{(1-\xp)+\om(1-\xi/\xp)}{\xp}\;,
\end{eqnarray}
while conservation of transverse momentum requires that $s_\perp$, $z_\perp$ and $\bar{z}_\perp$ satisfy the triangle inequalities
\begin{eqnarray}
\vert z_\perp-s_\perp\vert &\le& \bar{z}_\perp \\
\vert \bar{z}_\perp-s_\perp\vert &\le& z_\perp\;.
\end{eqnarray}

An additional variable, $\beta$, is required to parametrise the
permitted values of $z_\perp$ and $\bar{z}_\perp$. Let us choose to
write:
\begin{eqnarray}
z_\perp^2 &=&
4\eta\left[a\bar{\eta}+b\eta+2\cos\beta\sqrt{ab\eta\bar{\eta}}\right]
\\
\bar{z}_\perp^2 &=& 4\bar{\eta}\left[a\eta+b\bar{\eta}-2\cos\beta\sqrt{ab\eta\bar{\eta}}\right]\;,
\end{eqnarray}
where
\beq
a =
\frac{(1-\xp)(\xp-\om)}{\xp^2\bigl(1-\om\xi/\xp^2\bigr)^2}\hspace{0.5in}
b =
\frac{\om(1-\xi/\xp)(1-\om\xi/\xp)}{\xp^2\bigl(1-\om\xi/\xp^2\bigr)^2}\;.
\eeq

Given $s_\perp$, $z_\perp$ and $\bar{z}_\perp$, the angles $\chi$ and
$\bar{\chi}$ are determined up to a sign. We may then choose $\eta$
and $\beta$ as the independent variables, with phase space
\beq
0\le\eta\le 1-\om\xi/\xp^2\ind 0\le\be\le\pi\;.
\eeq

To integrate the matrix elements we apply the operator
\begin{equation}
\int\frac{d^3\bd{r}}{(2\pi)^3 2r^0}\frac{d^3\bd{k}}{(2\pi)^3 2k^0}\:(2\pi)^4\delta^4(q+s-r-k)\;.
\end{equation}
Integrating with respect to $\bd{k}$ and making substitutions for $\bd{r}$ gives
\begin{equation}
\frac{1}{4(2\pi)^2}\int\frac{z_\perp dz_\perp d\chi d\eta}{\eta}\:\frac{\delta\left(A-\sqrt{B-2s_\perp z_\perp\cos\chi}\right)}{\sqrt{B-2s_\perp z_\perp\cos\chi}}\;,
\end{equation}
where $A$ and $B$ do not depend on $\chi$.

Next we integrate over $\chi$. There are two values satisfying the integration condition, and they differ by a sign. This gives
\begin{equation}
\frac{1}{4(2\pi)^2}\int\frac{d\eta}{\eta}\frac{dz_\perp^2}{s_\perp z_\perp\vert\sin\chi\vert}\;.
\end{equation}

Applying the parametrisation in terms of $\beta$, we find that
\begin{eqnarray}
\left\vert\frac{\partial z_\perp^2}{\partial\beta}\right\vert &=& 8\eta\sin\beta\sqrt{ab\eta\bar{\eta}} \\
s_\perp z_\perp\vert\sin\chi\vert &=& 4(1-\om\xi/\xp^2)\sin\beta\sqrt{ab\eta\bar{\eta}}\;,
\end{eqnarray}
and therefore the integral operator for the photon-gluon fusion is
\begin{equation}
\frac{1}{8\pi^2(1-\om\xi/\xp^2)}\int_0^{1-\om\xi/\xp^2}d\eta\int_0^\pi d\beta\;.
\end{equation}

We calculate the quantities $C^{ij}(\xi,\xp,\om)$ as defined by
\beq
\label{Cij}
\frac{\as T_R}{2\pi}\left(\sum_{q^\prime}e_{q^\prime}^2\right)C^{ij}(\xi,\xp,\om) = \int_0^{2\pi}\frac{d\theta}{2\pi}F_{ij}(\xi,\xp,\om,\theta)\;,
\eeq
where the sum over $q^\prime$ represents the outgoing quark-antiquark flavours.
We first perform the $\theta$ integral: the contributions to
$F_L$ have no $\theta$-dependence, while the contributions to $F_T$
have terms proportional to $\cos\theta$ and $\cos^2\theta$. Then the
integration over $\beta$ can be performed by writing
$t=\tan(\beta/2)$, and that over $\eta$ by carefully expanding in
powers of $\om=\rho\xp^2$ as described in \cite{smye}. We obtain, for the
longitudinal structure function,
\begin{eqnarray}
\label{CLstart}
C_L^{11} &=& 32(1-\xp)\Bigl[(3+\xi-4\xp)+(1-\xp)\log\om\Bigr]\om +
\cl{O}(\om^2) \\ C_L^{12} = C_L^{21} &=&
-32i\sqrt{1-\xi/\xp}(1-\xp)(1-3\xp-\xp\log\om)\om + \cl{O}(\om^2)
\\ C_L^{22} &=& 8\xp(1-\xp) +
8\Bigl[(1/\xp+2\xi/\xp-10-8\xi+23\xp+9\xi\xp-17\xp^2) \nln & & +
(\xi/\xp-3-2\xi+8\xp+2\xi\xp-6\xp^2)\log\om\Bigr]\om +
\cl{O}(\om^2) \\ C_L^{33} &=& 8\xp(1-\xp) +
8\Bigl[(1/\xp+2\xi/\xp-10-8\xi+23\xp+7\xi\xp-15\xp^2) \nln & & +
(\xi/\xp-3-2\xi+8\xp+2\xi\xp-6\xp^2)\log\om\Bigr]\om +
\cl{O}(\om^2)\;.
\end{eqnarray}
Hence in the limit $\om\to 0$ we recover the well-known results
$C_L^{11}=C_L^{12}=C_L^{21}=0$ and $C_L^{22}=C_L^{33}=8\xp(1-\xp)$.

For the transverse structure function we obtain
\begin{eqnarray}
C_T^{11} &=& -8\xp(1-\xp) +
8\Bigl[(-2/3\xp+25/3+2\xi-15\xp-2\xi\xp+8\xp^2) \nln & & +
2(1-\xp)^2\log\om\Bigr]\om + \cl{O}(\om^2) \\ C_T^{12} = C_T^{21}
&=& \frac{-4i}{\sqrt{1-\xi/\xp}}\Bigl[(\xi/3\xp^2
-2/3\xp-11\xi/3\xp+10/3+18\xi-20\xp \nln & & -14\xi\xp+16\xp^2) -
2(1-\xp)(4\xp-3\xi)\log\om\Bigr]\om + \cl{O}(\om^2) \\ C_T^{22} &=&
-2(1-2\xp+2\xp^2)(2+\log\om) -
\frac{2}{1-\xi/\xp}\Bigl[(\xi^2/3\xp^3+8\xi/3\xp^2-8/3\xp \nln & &
+13\xi^2/3\xp^2-88\xi/3\xp+76/3-14\xi^2/\xp+60\xi-44\xp+16\xi^2-48\xi\xp
\nln & & +30\xp^2) +
2(\xi/\xp^2-1/\xp+\xi^2/\xp^2-7\xi/\xp+6-\xi^2/\xp+9\xi-7\xp \nln & &
+\xi^2-6\xi\xp+4\xp^2)\log\om\Bigr]\om + \cl{O}(\om^2) \\ C_T^{33}
&=& -2(1-2\xp+2\xp^2)(2+\log\om) -
\frac{2}{1-\xi/\xp}\Bigl[(-\xi^2/3\xp^3+4\xi/\xp^2-4/\xp \nln & &
+11\xi^2/3\xp^2-28\xi/\xp+24-18\xi^2/\xp+68\xi-52\xp+16\xi^2-48\xi\xp+34\xp^2)
\nln & & +
2(\xi/\xp^2-1/\xp+\xi^2/\xp^2-7\xi/\xp+6-3\xi^2/\xp+13\xi-11\xp \nln &
& +3\xi^2-10\xi\xp+8\xp^2)\log\om\Bigr]\om + \cl{O}(\om^2)\;.
\label{CTfinish}
\end{eqnarray}
As $\om\to 0$ the pieces $C_T^{22}$ and $C_T^{33}$ diverge
logarithmically: this is the piece associated with the collinear
splitting of the gluon into a quark-antiquark pair.

\sect{Power corrections}
\label{secglev}
We noted above that the interaction of the produced
virtual gluon is given independently of the production mechanism by
the function $F_{ij}(\xi,\xp,\om,\theta)$, where $i$ and $j$
represent polarisations and $\xi$, $\xp$, $\om$ and $\theta$
parametrise the gluon's 4-momentum. In general, calculations of power
corrections will depend on all four of these quantities.

Thus a natural and intuitive way to combine this with a gluon distribution function might be to write
\beq
\label{gijconv}
F(x) =
\sum_{i,j}\int_x^1\frac{d\xi}{\xi}\int_\xi^1\frac{d\xp}{\xp}\int_0^{2\pi}\frac{d\theta}{2\pi}\int_0^\xp\frac{d\om}{\om}\;g_{ij}(x/\xp,\om,\xi/\xp)F_{ij}(\xi,\xp,\om,\theta)\;,
\eeq
where $g_{ij}(x/\xp,\om,\xi/\xp)$ is the relevant gluon distribution function
which by symmetry does not depend on $\theta$. Of course this needs to
be treated with some care, since both halves of the
diagram in figure \ref{decompfig} are gauge-dependent and so therefore are
all these distributions when the gluon is off-shell, i.e.~$\om\ne 0$.
Nevertheless, we can still make progress: we present an intuitive argument
that, although mathematically not totally rigorous, is helpful in understanding
the underlying physics.

For the diagonal elements we can view this probabilistically: the
expected number of gluons of polarisation $i$ in an element of
parameter space $du\,d\om\,dz$ at $(u,\om,z)$, where $u=x/\xp$ is the 
longitudinal momentum fraction of the proton carried by the gluon, and
$z=\xi/\xp$, is:
\beq
g_{ii}(u,\om,z)du\frac{d\om}{\om}\frac{dz}{z}\;.
\eeq

This is intended to be schematic only\footnote{It 
could of course be more rigorously formulated in terms of
well-defined unintegrated parton distributions, but that is unnecessary for
the purposes of the argument given here.} --- clearly any gluon not
collinear with the proton, i.e.~with $\om\neq 0$, will experience the
confining effect of the QCD potential, and so cannot exist for more
than a short time (where `short' in this context means $\sim 
1/\Lambda_\sml{QCD}$). Such particles are not asymptotically free: they
can only be resolved at high momentum
scales, and it is precisely this resolution of gluons with non-zero
transverse momentum that gives rise to the running of the parton
distributions. Power corrections are also known to arise from the 
interactions of
gluons with non-zero transverse momentum (see e.g.~\cite{efp}), and so are
also generated by the resolution of short-lived virtual gluons within the
proton. Thus we expect power corrections in singlet DIS to be
intimately related to the running of parton distributions.

Consider first the transverse polarisations: these are the physical
polarisations. Speaking schematically (i.e.~not being too precise about
our choice of factorisation scheme), we may view the polarised
gluon distribution function $g_i(u,M^2)$ as representing the gluons with
transverse momentum less than $M$. So
\beq
g_i(u,M^2) =
\int_{u}^{1}\frac{dz}{z}\int_0^{\bar{\om}(M^2)}\frac{d\om}{\om}g_{ii}(u,\om,z)\;,
\eeq
for $i=2,3$, where $\bar{\om}(M^2)$ is the value of $\om$
corresponding to a transverse momentum $M^2$. Therefore,
\beq
\label{intgii}
M^2\frac{\partial g_i(u,M^2)}{\partial M^2} =
\int_{u}^{1}\frac{dz}{z}\;\frac{M^2}{\bar{\om}}\frac{\partial\bar{\om}}{\partial
M^2}g_{ii}(u,\bar{\om},z)\;,
\eeq
and, using equation \rf{sperp}, we obtain
\beq
\frac{M^2}{\bar{\om}}\frac{\partial\bar{\om}}{\partial M^2} = \frac{1-\bar{\om}z}{1-2\bar{\om}z} = 1 + \cl{O}(\bar{\om})\;.
\eeq
Further, we know from the DGLAP equation that
\beq
M^2\frac{\partial g_i(u,M^2)}{\partial M^2} =
\frac{\as}{2\pi}\int_{u}^1\frac{dz}{z}\left[C_Fq_\sml{tot}(u/z)P_{q\to
g}^i(z) + C_Ag(u/z)P_{g\to g}^i(z)\right]\;,
\eeq
so for the sake of illustration let us make the simplest consistent assignment for the differential gluon distribution $g_{ii}(u,\bar{\om},z)$, which is
\beq
\label{gapprox}
g_{ii}(u,\bar{\om},z) = \frac{\as}{2\pi}\left[C_Fq_\sml{tot}(u)P_{q\to
g}^i(z) + C_Ag(u)P_{g\to g}^i(z)\right]+\cl{O}(\bar{\om})\;.
\eeq
Here $q_\sml{tot}(u)=\sum_q[q(u)+\bar{q}(u)]$ is the total quark and antiquark content of the proton.

So let us return to \rf{gijconv} and define the coefficient function
\beq
C_g(\xp) = F_{ii}(\xi,\xp,0,\theta)
\eeq
which is independent of $\xi,\theta$ and polarisation $i$. Let us separate off the piece that generates the collinear divergence by writing
\beq
\int_0^{2\pi}F_{ii}(\xi,\xp,\om,\theta)\frac{d\theta}{2\pi} = C_g(\xp)\Theta(\bar{\om}(M^2)-\om) + \delta F_{ii}(\xi,\xp,\om) \;;
\eeq
we then obtain the contribution to the observable
$F$ from incoming transversely-polarised gluons as:
\begin{eqnarray}
F_\sml{trans}(x) &=&
\sum_{i=2}^3\int_x^1\frac{d\xi}{\xi}\int_\xi^1\frac{d\xp}{\xp}\int_0^\xp\frac{d\om}{\om}\int_0^{2\pi}\frac{d\ta}{2\pi}\;g_{ii}(x/\xp,\om,\xi/\xp)F_{ii}(\xi,\xp,\om,\ta)\\
\label{factored}
&=& \int_x^1\frac{d\xp}{\xp}g(x/\xp,M^2)C_g(\xp) + \nln
& & \ind\sum_{i=2}^3\int_x^1\frac{d\xi}{\xi}\int_\xi^1\frac{d\xp}{\xp}\int_0^\xp\frac{d\om}{\om}\;g_{ii}(x/\xp,\om,\xi/\xp)\delta F_{ii}(\xi,\xp,\om)\;.\ind
\end{eqnarray}

The first term on the right hand side of \rf{factored} is simply the
standard contribution to photon-gluon fusion, evaluated with an
on-shell initial gluon and convoluted with the gluon
distribution $g(x)=\sum_{i=2}^3 g_i(x)$. This contains logarithmic
scaling violations given by
the running of the gluon distribution, but contains no power
corrections associated with the virtuality of this gluon.

The second term gives us the required power corrections, since this is the term with an integral over the gluon's virtuality. Let us now substitute the differential gluon distribution with expression \rf{gapprox}, thus obtaining
\bea
\label{schematic}
\lefteqn{\delta F_\sml{trans}(x) = \sum_{i=2}^{3}\int_x^1\frac{d\xi}{\xi}\int_\xi^1\frac{d\xp}{\xp}\int_0^\xp\frac{d\om}{\om}g_{ii}(x/\xp,\om,\xi/\xp)\delta F_{ii}(\xi,\xp,\om)}\\
&=& \!\frac{\as}{2\pi}\!\sum_{i=2}^{3}\int_x^1\!\frac{d\xi}{\xi}\!\int_\xi^1\!\frac{d\xp}{\xp}\!\int_0^\xp\!\frac{d\om}{\om}\!\left[C_Fq_\sml{tot}(x/\xi)P_{q\to g}^i(\xi/\xp)+C_Ag(x/\xi)P_{g\to g}^i(\xi/\xp)+\cl{O}(\om)\right]\delta F_{ii}\,.\nonumber
\eea

Although the above argument has been restricted to the contributions from
transverse polarisations, and even these have been treated only schematically,
we are now in a position to be able to see what is happening physically. We
have gone round in a big circle: starting with figures \ref{splitfig}
and \ref{spglufig}, we have detached the lower parts of the diagrams
corresponding to gluon production, to give the piece corresponding to
photon-gluon fusion. But then in order to convolute with the correct
differential gluon distribution within the proton we had to make use
of the DGLAP equation, which reintroduced those lower legs of the
diagrams. Thus for the purposes of power corrections we find that the
lower parts of those diagrams genuinely are important and cannot be removed.
This also explains the presence of the longitudinal gluon polarisation:
we are not dealing with a real incoming gluon and so we need not be
restricted to physical polarisations. Equation \rf{schematic} is intended
to be illustrative only --- it was not rigorously derived. Yet it is clear that
while the leading perturbative contribution to singlet DIS, given by 
the first term in
equation \rf{factored}, is $\cl{O}(\as)$, the $1/Q^2$ power corrections as
well as the logarithmic scaling violations are $\cl{O}(\as^2)$, the 
additional factor of $\as$ arising from the DGLAP equation. (Contrast
this with flavour non-singlet DIS, whose lowest order is
$\cl{O}(\as^0)$ but which has leading power corrections at $\cl{O}(\as)$.)

Thus the leading singlet power corrections are from the diagrams \ref{splitfig} and \ref{spglufig} as they appear: we have two independent renormalon chains, and the magnitude of the power corrections is given by the quantities $D_1$ and $D_2$ of equations \rf{D1def} and \rf{D2def}.

The characteristic function $\hat{\cl{F}}(\e)$ generating power corrections to structure functions is thus given by
\beq
\hat{\cl{F}}(\eps) = \left(\sum_{q^\prime}e_{q^\prime}^2\right)\int_x^1\frac{d\xi}{\xi}\left\{\frac{T_R C_F}{(2\pi)^2}q_\sml{tot}(x/\xi)C^q(\xi,\eps) + \frac{T_R C_A}{(2\pi)^2}g(x/\xi)C^g(\xi,\eps)\right\}\;,
\eeq
where
\beq
C^{q/g}(\xi,\eps) =
\int_\xi^1\frac{d\xp}{\xp}\int_0^{1/\xp}\frac{d\rho}{\rho+\eps}\sum_{i,j}M_{ij}^{q/g}(\xi/\xp,\rho\xp^2)C^{ij}(\xi,\xp,\rho\xp^2)\;,
\eeq
and the $C^{ij}$ were defined in equation \rf{Cij} and calculated in \rf{CLstart} to \rf{CTfinish}. The sum over $q^\prime$ represents outgoing quark-antiquark flavours in the photon-gluon fusion. We retain terms in $\hat{\cl{F}}(\eps)$ up to $\cl{O}(\eps)$ that are non-analytic as $\eps\to 0$. In order to define a gauge-invariant quantity with a natural scale for the running coupling, we use the matrix \rf{Mpt} evaluated using the pinch technique --- the remaining diagrams are also expected to give power corrections, but as in \cite{cpar} they are expected to have a different structure.

Using \rf{split2} and \rf{glusp2} we find
\begin{eqnarray}
\xi C_T^q &=& -{\textstyle\frac{2}{9}}(2-63\xi+63\xi^2-2\xi^3+12\log\xi-27\xi\log\xi-27\xi^2\log\xi+12\xi^3\log\xi)\log\eps\nln
& & +{\textstyle\frac{2}{3}}(4+3\xi-3\xi^2-4\xi^3+6\xi\log\xi+6\xi^2\log\xi)(\log\xi-1+\half\log\eps)\log\eps\nln
& & +{\textstyle\frac{2}{5}}(2+25\xi^2-25\xi^3-2\xi^5+15\xi^2\log\xi+15\xi^3\log\xi)\eps\log\eps\nln
& & -2(5\xi^2-5\xi^3+2\xi^2\log\xi+2\xi^3\log\xi)(\log\xi-1+\half\log\eps)\eps\log\eps\;,\\
\xi C_L^q &=& -{\textstyle\frac{8}{3}}(1-3\xi+2\xi^3-3\xi^2\log\xi)\log\eps\nln
& & -{\textstyle\frac{8}{225}}(17+75\xi^2-125\xi^3+33\xi^5+30\log\xi+75\xi^3\log\xi-45\xi^5\log\xi)\eps\log\eps\nln
& & +{\textstyle\frac{8}{15}}(2-15\xi^2+10\xi^3+3\xi^5-15\xi^3\log\xi)(\log\xi-1+\half\log\eps)\eps\log\eps\;,
\end{eqnarray}
which is the result given in \cite{smye}, and
\begin{eqnarray}
\xi C_T^g &=& -{\textstyle\frac{4}{9}}(1-99\xi+99\xi^2-\xi^3+6\log\xi-27\xi\log\xi-54\xi^2\log\xi+33\xi^3\log\xi)\log\eps\nln
& & +{\textstyle\frac{4}{3}}(2+6\xi+3\xi^2-11\xi^3+3\xi\log\xi+12\xi^2\log\xi)(\log\xi-1+\half\log\eps)\log\eps\nln
& & +{\textstyle\frac{4}{5}}(1+40\xi^2-35\xi^3-6\xi^5+15\xi^2\log\xi+30\xi^3\log\xi)\eps\log\eps\nln
& & -8(5\xi^2-5\xi^3+2\xi^2\log\xi+2\xi^3\log\xi)(\log\xi-1+\half\log\eps)\eps\log\eps\;,\\
\xi C_L^g &=& -{\textstyle\frac{8}{3}}(1-3\xi-9\xi^2+11\xi^3-12\xi^2\log\xi)\log\eps\nln
& & -{\textstyle\frac{8}{225}}(17+1275\xi^2-1475\xi^3+183\xi^5+30\log\xi+450\xi^2\log\xi\nln
& & \hspace{2.0cm}+750\xi^3\log\xi-270\xi^5\log\xi)\eps\log\eps\nln
& & +{\textstyle\frac{16}{15}}(1-15\xi^2+5\xi^3+9\xi^5-30\xi^3\log\xi)(\log\xi-1+\half\log\eps)\eps\log\eps\;.
\end{eqnarray}

Thus we obtain the $1/Q^2$ power corrections to $F_T(x)$ and $F_L(x)$: those arising from the quark distribution are
\beq
\delta F_i^q(x) = \frac{T_RC_F}{(2\pi)^2}\left(\sum_{q^\prime}e_{q^\prime}^2\right)\int_x^1\frac{d\xi}{\xi}q_\sml{tot}(x/\xi)\delta C_i^q(\xi)
\eeq
where
\bea
\delta C_T^q(\xi) &=& \frac{D_1}{Q^2}\biggl[\frac{2(2+25\xi^2-25\xi^3-2\xi^5+15\xi^2\log\xi+15\xi^3\log\xi)}{5\xi}\nln
& & \qquad-2(5\xi-5\xi^2+2\xi\log\xi+2\xi^2\log\xi)\log\frac{D_2\xi}{eQ^2}\biggr]\\
\delta C_L^q(\xi) &=& \frac{D_1}{Q^2}\biggl[-\frac{8(17+75\xi^2-125\xi^3+33\xi^5+30\log\xi+75\xi^3\log\xi-45\xi^5\log\xi)}{225\xi}\nln
& & \qquad+\frac{8(2-15\xi^2+10\xi^3+3\xi^5-15\xi^3\log\xi)}{15\xi}\log\frac{D_2\xi}{eQ^2}\biggr]\;;
\eea
and those from the gluon distribution are
\beq
\delta F_i^g(x) = \frac{T_RC_A}{(2\pi)^2}\left(\sum_{q^\prime}e_{q^\prime}^2\right)\int_x^1\frac{d\xi}{\xi}g(x/\xi)\delta C_i^g(\xi)
\eeq
where
\bea
\delta C_T^g(\xi) &=& \frac{D_1}{Q^2}\biggl[\frac{4(1+40\xi^2-35\xi^3-6\xi^5+15\xi^2\log\xi+30\xi^3\log\xi)}{5\xi}\nln
& & \qquad-8(5\xi-5\xi^2+2\xi\log\xi+2\xi^2\log\xi)\log\frac{D_2\xi}{eQ^2}\biggr]\\
\delta C_L^g(\xi) &=& \frac{D_1}{Q^2}\biggl[-\frac{8}{225\xi}\left(\parbox{3.0in}{$17+1275\xi^2-1475\xi^3+183\xi^5+30\log\xi+450\xi^2\log\xi+750\xi^3\log\xi-270\xi^5\log\xi$}\right)\nln
& & \qquad+\frac{16(1-15\xi^2+5\xi^3+9\xi^5-30\xi^3\log\xi)}{15\xi}\log\frac{D_2\xi}{eQ^2}\biggr]\;.
\eea

\begin{figure}[ht]
\begin{center}
\epsfig{file=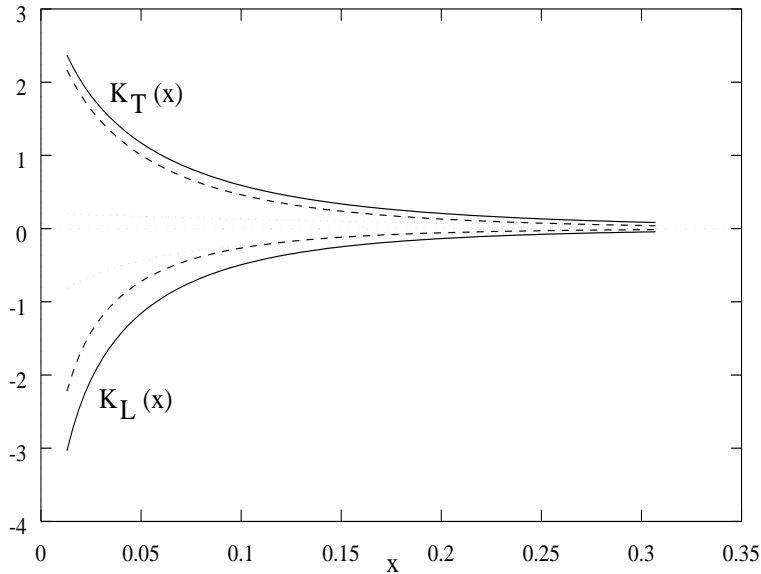, height=3.0in, width=4.0in}
\caption{\label{graphtl}Graph showing $K_T(x)$ and $K_L(x)$. The dotted lines represent the contributions from the quark distribution; the dashed lines represent the gluon contributions. The totals are represented by the solid lines.}
\end{center}
\end{figure}

\sect{Results and conclusions}
\label{conc}
Figure \ref{graphtl} shows the magnitudes $K$ of the $1/Q^2$ power corrections to structure functions, given by
\beq
\frac{\delta F_{T/L}^{q/g}(x)}{F_T^{(0)}(x)} = \frac{D_1}{Q^2}K_{T/L}^{q/g}(x)\;,
\eeq
where $F_T^{(0)}(x)$ is the Born-level result for the transverse structure function given in equation \rf{bornsf}.
These were calculated at $Q^2 = 500\mbox{ GeV}^2$, using the
corresponding MRST (central gluon) parton distributions \cite{mrst},
and assuming four flavours of outgoing quark-antiquark pairs. The
unknown value of $D_2/e$ was set to $0.06\mbox{ GeV}^2$,
i.e.~approximately $\Lambda^2$, although the qualitative behaviour of
these power corrections does not change provided we keep $D_2\ll Q^2$.

The dominant contribution to both of these comes from the gluon
distribution, as may be expected at these relatively low values of
$x$. All the contributions tend to zero at large $x$, and become large at small
$x$. The contributions to $K_T$ are positive and those to $K_L$
negative, but they have similar magnitudes.

\begin{figure}[ht]
\begin{center}
\epsfig{file=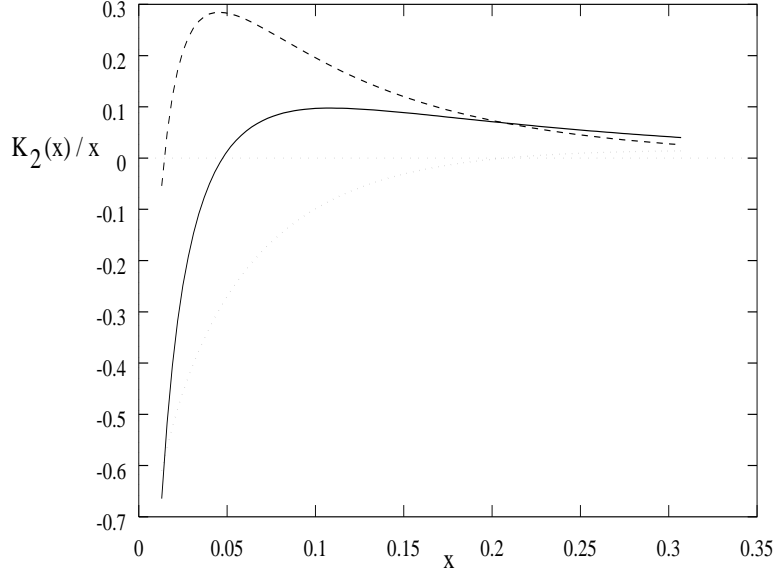, height=3.0in, width=4.0in}
\caption{\label{graph2}Graph showing $K_2(x)/x$. The dotted line represents the contribution from the quark distribution; the dashed line represents the gluon contribution. The total is represented by the solid line.}
\end{center}
\end{figure}

The corresponding quantity related to $F_2/x$, which is $K_2/x = K_T +
K_L$, is shown in figure \ref{graph2}. The positive and negative
contributions partially cancel, giving a power correction smaller by a
factor 3 or 4 than that for $F_T$ or $F_L$.

The power corrections are multiplied by the unknown factor $D_1$,
defined in \rf{D1def}. If $D_1$ is positive, we might expect $F_T(x)$
to show a positive $1/Q^2$ power correction and $F_L(x)$ a negative
one. These results are all qualitatively the same as those in \cite{smye},
which took into account only the quark contribution, 
but the inclusion of the gluon contribution increases
considerably the size of the correction, in the case of $F_T$ by more
than an order of magnitude.

\sect{Discussion}
The calculation of power corrections using the renormalon model or the dispersive approach in flavour singlet DIS is a non-trivial problem, due to the fact that figure \ref{decompfig} as it stands is gauge dependent, both in the upper and lower halves of the diagram. These halves may both be considered to be cut insertions into a single gluon propagator, albeit in another channel, and so the pinch technique can be used to define a natural gauge-invariant quantity. The diagram \ref{decompfig} evaluated using the pinch technique gives only one gauge-independent part of the full $\cl{O}(\as^2)$ amplitude, and the expectation is that the other parts will also give power corrections, although not taking the form of two renormalon chains with equal 4-momenta.

Power corrections thus arise when a virtual gluon is radiated from a quark or gluon, and this virtual gluon then interacts with the photon. We found that, as well as the physical transverse polarisations, we also have contributions to power corrections from longitudinal polarisations and interference terms. In addition, since the gluon's 4-momentum is not simply some multiple of that of the proton, the contribution from the photon-gluon fusion is a function of all four momentum components. In particular the kinematics of the virtual gluon production are important: where it is radiated from an on-shell parton, the gluon remembers the momentum fraction of the parton that emitted it. This, along with the discussion of the virtual gluon distribution in section \ref{secglev}, indicates that we cannot detach the two halves of figure \ref{decompfig} but rather that the power corrections arise from the full diagram and are thus an $\cl{O}(\as^2)$ effect.

Calculations of power corrections to structure functions and fragmentation functions due to gluon radiation by a quark are already published in the literature \cite{stein,smye,smye2}. However the contributions from radiation by a gluon dominate, because of the relative behaviours of the quark and gluon parton densities. This larger correction was evaluated above for the structure functions. Above $x=0.05$ the power corrections to the structure functions are small, but we predict that for $x$ below $0.05$ the corrections grow. So the correction to $F_2$ at $x=0.01$ and $Q^2=4$ GeV$^2$ might be around --2\% (assuming a reasonable value of $D_1\approx 0.1$ GeV$^2$). The corrections to $F_T$ and $F_L$ are expected to be slightly larger. In any case, we do not expect to see the large ($\sim 50\%$) corrections predicted by \cite{stein}.

It was seen that the use of the pinch technique lead to a failure to recover the usual $g\to g$ splitting function. This is not a problem --- the remaining pieces come from the remaining diagrams. All the pinch technique has done is move certain terms from one diagram to another. But the point of using this particular separation of terms is that one can define a gauge-invariant QCD effective charge, so the virtuality of the gluon enters as the natural scale for the coupling; it is not clear that this is the case for the other terms.

Finally, there is clearly some similarity between the situation here and that of a decaying outgoing virtual gluon in $\ee$ annihilation (figure \ref{eefig}), where we also have two renormalon chains with equal 4-momenta $k$, for some $k$. Again we have an integral over the virtuality $k^2$, in which the coupling appears as $\vert\as(-k^2)\vert^2$. But in that case we use equation \rf{rhosdef} to convert $\as^2$ into the spectral function $\rho_s(k^2)$, and the two chains became one, with a cut bubble insertion. There are two reasons why we cannot do the same here. Firstly, in the $\ee$ case, the virtual gluon is timelike, $k^2$ is positive and thus we have $\rho_s(k^2)$ integrated over positive $k^2$. But here the gluon is spacelike, so $k^2$ is negative, and while it is quite possible to convert the two factors of $\as$ into a single $\rho_s(k^2)$, it is integrated over negative values of its argument. It is therefore not naturally manipulated into standard single-chain form. Secondly, although the algebra of the Feynman diagrams is identical (i.e.~related by crossing), in the $\ee$ case we can simply integrate over the cut bubble but in DIS we must include parton density functions for the incoming quark or gluon. This makes it quite impossible to integrate out the crossed cut bubble that is the mechanism for production of the virtual gluon. So, while there are interesting and useful parallels between power corrections to singlet DIS and $\ee$ annihilation with outgoing gluon splitting, there are also significant differences and the relationship between them is not as simple as may naively have been supposed.

\section*{Acknowledgements}
The author wishes to thank Yuri Dokshitzer, Mark Smith, David Summers, Jay Watson and Bryan Webber for helpful discussions and comments.

\end{document}